\documentclass[aps,pra,showpacs,preprint,amsmath,amssymb]{revtex4}

\usepackage{graphicx}
\usepackage{dcolumn}
\usepackage{bm}
\usepackage{xcolor,soul}
\newcommand{\beq}{\begin{equation}}
\newcommand{\eeq}{\end{equation}}

\sethlcolor{yellow}
\setulcolor{red}

\begin{document}

\title{Ion dynamics  in a linear radio-frequency trap\\
 with a single cooling laser}

\author{M. Marciante}
 \email{mathieu.marciante@etu.univ-provence.fr}
\author{C. Champenois}
\author{A. Calisti}
\author{J. Pedregosa-Gutierrez}
\author{M. Knoop}

\affiliation{Physique
des Interactions Ioniques et Mol\'eculaires, UMR 6633 CNRS et Aix-Marseille Universit\'e,
 Centre de Saint J\'er\^ome, Case C21,
13397 Marseille Cedex 20, France}%

\date{\today}

\begin{abstract}
We analyse the possibility of  cooling ions with a single laser beam, due to the coupling between the three
components of their motion induced by the Coulomb interaction. For this purpose, we numerically study  the
dynamics of  ion clouds of up to 140 particles, trapped in a linear quadrupole potential and  cooled with a
laser beam propagating in the radial plane. We use Molecular Dynamics simulations and model the laser cooling
by a stochastic process.  For each component  of the motion, we systematically study  the dependence of the
temperature  with the anisotropy  of the trapping potential.
Results obtained using the full  radio-frequency (rf) potential are compared to those of the corresponding
pseudo-potential. In the rf case, the rotation symmetry of the potential has to be broken to keep ions inside
the trap. Then, as for the pseudo-potential case, we show that the efficiency  of the Coulomb coupling to
thermalize the components of motion depends on the geometrical configuration of the cloud. Coulomb coupling
appears to be not efficient when the ions organise as a line or a pancake and the three components of motion
reach the same temperature only if the cloud extends in three dimensions.
\end{abstract}


\maketitle

\section{\label{sec:Intro}Introduction}

The advent of laser cooling \cite{hansch75,wineland75} generalized infinite trapping times in radio-frequency
(rf) traps \cite{paul90}  and allowed observing a single atom by its emitted fluorescence \cite{neuhauser80}.
Doppler cooling in rf traps only employs a single laser beam, as the trapping potential brings back the atom
into resonance when it is counter-propagating with the laser beam. Although heating is expected perpendicularly  to the laser beam from the
mechanical effect of light \cite{wineland87}, several experiments with
single ions in spherical Paul traps have demonstrated cooling to the Doppler limit with a single laser beam, as
long as it does not propagate along the symmetry axis of the trap \cite{bergquist87,diedrich89,monroe95}. This
is attributed to trap defects which induce a coupling between the three components of the motion
\cite{wineland79,bergquist87}. For ions in a linear quadrupole trap, the use of a single laser beam to
completely cool a trapped sample in its three degrees of freedom remains an open question which we address by
numerical simulations in the present work. Coulomb repulsion between the ions is expected to couple their
motion and to result in the thermalization of the three degrees of freedom, independently of the direction of
the laser beam. Nevertheless, depending on the aspect ratio of the cloud, the coupling may be more or less
efficient. The dependance of this aspect ratio  with the anisotropy of the confining pseudo-potential, as well
as the transitions between different organized structures have been studied numerically by J. Schiffer in
\cite{schiffer93} and analytically by D. Dubin \cite{dubin93b}. These works show that the minimal energy
configurations can be one, two or three dimensional, depending on the aspect ratio of the potential and the
number of trapped ions. In the present work, we numerically study  the thermalization of the different degrees of
motion under Doppler laser cooling by Molecular Dynamics methods. The study is carried out for different aspect
ratios and  ion numbers, in order to determine the influence of the cloud anisotropy on the cooling of the
sample. In the simulations, Doppler laser cooling is introduced by the mechanical effect of  absorption and
emission of photons by the atoms.

Laser cooling of an ion string has been previously studied analytically using the common vibrational mode to describe the motion of the ions \cite{morigi99,morigi01,morigi03b}. In these works, the string is assumed to be tightly bound in the radial plane and the only considered motion is the one along the trap axis. Contrary to our configuration, this motion is directly laser cooled and its thermalization with the motion in the radial plane is not addressed in these works. Studies on ion strings \cite{morigi03b} and more generally on ion crystals \cite{javanainen86} have shown that the motion along one direction is cooled provided  the cooling laser  has a projection along it. The role of the rf-driven motion on the cooling efficiency has been analyzed for crystals \cite{devoe89} as well as for single ions \cite{cirac94}, pointing the influence of the laser detuning compared to the rf frequency.

The present article is organized as follows, we first introduce the studied system and the numerical method used. Then,
in the following section, we study the effect of the anisotropy  of the cloud on the cooling, in the adiabatic
approximation where the rf trapping field is modeled by a static pseudo-potential. In the last section, the same
properties are analyzed in the case where the full rf electric field is introduced to trap the ions.

\section{\label{sec:Framework}Framework}
\begin{figure}
    \includegraphics[width=8cm]{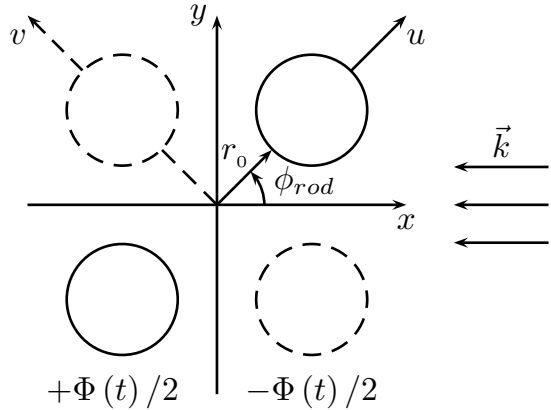}
    \caption{Geometry of the trap  in the $\left(x,y\right)$-plane. $\vec{k}$ is the laser wave-vector. A voltage $\pm \Phi(t)/2$ is applied on each pair of opposite rods.}
    \label{fig_trap}
\end{figure}

The system under study is an ensemble of $N$   ions of mass $m$ and charge $q_e=+e$, confined in a linear
quadrupole trap and Doppler laser cooled by  a single laser beam of wave-vector $\vec k$. The trapping
field is represented either by the time dependent electric field (the rf-case) or by the gradient deduced from
the static pseudo-potential (pseudo-potential case), associated to the rf electric field in the  adiabatic
approximation \cite{dehmelt67,paul90}.
We assume that the  laser beam intensity is  uniform over the area covered by the ions. The $x$ axis is always
defined by the propagation of the cooling laser beam and a $(u,v)$ axis system is related to the trap electrodes. An eventual trap rotation  in the $(x,y)$ plane with respect to the laser beam is described by  the  angle $\phi_{rod}$ (see Fig.~\ref{fig_trap}).
\subsection{\label{sec:dynamics}Dynamics}
In this section, we explicitly  write the equations for the forces acting on each ion. They result from the
trapping potential, the Coulomb repulsion between ions and the absorption-emission process implied in the
Doppler laser cooling. The rf electric field assures trapping in the radial plane of the trap. To achieve
trapping along the symmetry  axis ($z$ axis), a static voltage is applied to additional electrodes  at both ends of the
trap. In the center of the trap, the created potential can be considered as harmonic \cite{pedregosa10} and
characterized by the oscillation frequency $\omega_z$. The full trapping potential $U_{trp}$ is:
\begin{eqnarray}
U_{trp} &=& \frac{q_e \Phi\left(t\right)}{2r_{_0}^2}\left[\left(x^2-y^2\right) \cos\left(2 \phi_{rod}\right) + 2 xy \sin\left(2
\phi_{rod}\right)\right] \nonumber \\
&& -\ \frac{\ {m\ \omega_z}^2}{2} \left(\frac{x^2+y^2}{2}-z^2\right)
\label{eq_rftrap}
\end{eqnarray}
with $r_{_0}$  the inner radius of the trap and $ \Phi\left(t\right) = V_{dc} - V_{_0} \cos\left(\Omega_{_{rf}}
t\right)$ the potential difference applied between neighboring electrodes. The motion of a single ion due to the
quadrupole electric field is characterized by the Mathieu parameters \cite{McLachlan47,gosh_book} which depend
on the experimental parameters like:
\begin{eqnarray}
    q_u &=& \frac{2\ q_e\ V_{_0}}{m\ r_{_0}^2\ \Omega_{_{rf}}^2}, \\
    a_u &=& \frac{4\ q_e\ V_{dc}}{m\ r_{_0}^2\ \Omega_{_{rf}}^2}.
\end{eqnarray}
The radial  contribution of the harmonic static potential  (see Eq.~\ref{eq_rftrap}) can be taken into account
by  an effective  Mathieu parameter \cite{drewsen00}, denoted $\Delta a$ and which can be expressed like
$\Delta a =2 \omega_z^2 /\Omega_{_{rf}}^2$. These parameters define the oscillation frequencies of an ion in the
radial plane of the trap. In the adiabatic approximation \cite{dehmelt67}, the dynamics of one ion is the superposition of a harmonic motion (called macro-motion) and the rf-driven motion (called micro-motion).  The harmonic contribution is derived from the pseudo-potential $\Psi_{trp}$,   expressed in the  frame $(u,v)$ as:
\beq
    \Psi_{trp}=\frac{1}{2}m\omega_{r+}^2 u^2 + \frac{1}{2}m\omega_{r-}^2 v^2 +\frac{1}{2}m\omega_{z}^2 z^2.
    \label{eq_pseudopot}
\eeq
For $(|a_u|, q_u^2/2) \ll 1$ (and in practice $q_u < 0.4$), the oscillation frequencies can be written as
\beq
    \omega_{r\pm}=\frac{\Omega_{_{rf}}}{2}\sqrt{\frac{q_u^2}{2}\pm a_u -\Delta a}.
\eeq
In the case where no static voltage $V_{dc}$ is applied to the quadrupole rods, $\omega_{r+}=\omega_{r-}$, and the
anisotropy of the potential can be characterized by the ratio of its  strengths \cite{schiffer93}. When the
cylindrical symmetry is broken by a static voltage $V_{dc}$  applied to the rf rods (for $V_{dc} > 0$,
$\omega_{r+} > \omega_{r-}$), it has been shown \cite{schiffer93} that the ion structure aligns along the
minimal steepness direction $v$. Therefore, we consider that $ \omega_{r-}$ becomes the relevant parameter to
describe  the anisotropy of the trap:
\beq
    \alpha = \left(\omega_z/\omega_{r-}\right)^2.
\eeq

The  total Coulomb repulsion on one ion $i$ at position $\vec r_i$ is  the sum over the  $(N-1)$ other ions of the
Coulomb inter-particle force:
\beq
    \vec F_{C_i}=\sum_{j \neq i} \frac{q_e^2}{4 \pi \varepsilon_{_0}}\ \frac{(\vec r_i - \vec r_j)}{|\vec r_i - \vec r_j|^3}.
\eeq

Doppler laser cooling is simulated by kicks on the velocity of each ion for each absorption and emission process
\cite{blumel88}. On absorption, the momentum transferred to the ion is the momentum of a photon from the laser
beam $\hbar \vec k = - \hbar k \hat x$ (the laser propagation being along $- x$, see Fig.~\ref{fig_trap}). The
velocity recoil associated to the emission of one photon is uniformly distributed on a sphere of radius $\hbar
k/m$.  For a single emission event, the direction of the recoil is randomly distributed by the use of equation
\beq
    \overrightarrow{\delta v_{em}}= \frac{\hbar\ k}{m}\ \left( \sqrt{\theta_1 (1-\theta_1)}\
    \left[ \cos\left(\theta_2\right) \hat x + \sin\left(\theta_2\right) \hat y\ \right]  +(1-2\ \theta_1) \hat z \right)
\eeq
where $\theta_1$ and $\theta_2$ are stochastic variables uniformly distributed between $0$ and $1$. To calculate
the probability of absorption on the cooling transition, the  ion is considered as a two level system,
characterized by  the spontaneous emission rate of the excited level $\Gamma_{_0}$, far larger than the recoil
frequency $\hbar k^2/(2m)$. The strength of the laser-atom coupling is defined by the Rabi frequency $\Omega_r$.
The relevant time scale of the laser-atom interaction is of the same order of magnitude as the integration time
step. Nevertheless, we neglect the time correlation effect between the motion and the laser-atom interaction and
assume that for each time step, the stationary internal state is reached.  In the frame of these assumptions,
the probability $P_{abs}(t)$  for an ion  in the ground state with velocity $\vec v(t)$ to absorb a photon
during the time step $\Delta t$   is given by
\beq
    P_{abs} = \frac{\Gamma_{_0} \Delta t}{2}\ \frac{\Omega_r^2/2}{\left(\delta_{_l}-\vec k\cdot \vec
    v(t)\right)^2+\left(\Gamma_{_0}/2\right)^2+\Omega_r^2/2}
\eeq
and only depends on the velocity of the ion by the Doppler shift, $\delta_{_l}$ being the laser detuning for
an ion at rest. The stimulated emission is neglected and the probability $P_{em}$ for an ion in the excited
state to spontaneously emit a photon during the time step $\Delta t$ follows the usual decay law $ P_{em} =
\Gamma_{_0} \Delta t$. This model for Doppler laser cooling implies that the numerical integration keeps track
of the internal state of each ion. At each time step, the $N$ ions can change their internal state. The process
is controlled individually  by the comparison between a uniform random number associated to a single ion and the
absorption probability if the ion is in the ground state or to the emission probability otherwise.

One of the parameters  characterizing  the evolution of the system is its temperature. As we are interested
in the coupling between the three components of motion, we do not assume a uniform temperature for the three
axes and define a temperature for each of them $T_k$, with $k=x, y, z$. As the rf driven motion (or micromotion)
does not contribute to the temperature \cite{schiffer03,schiller03,zhang07}, our definition of the temperature
uses the part of the velocity which does not show the rf periodicity:
\begin{equation}
    T_k = \frac{1}{k_{_B}}\ m\left( \left< \overline{v^{2}}_k \right> - \left< \overline{v}_k \right>^2 \right)
    \label{eq_T}
\end{equation}
where the overline holds for the average over one rf period. The $<>$-brackets are the statistical averages over
the number of ions. This definition also subtracts  a possible motion of the center of mass (COM).

\subsection{\label{sec:Numerical}Numerical issues}
The time evolution of a system made of $N=35, 70$ and 140 trapped ions is studied  via Molecular Dynamics
simulations. We use the velocity-Verlet algorithm for the  integration of the equations of motion and the time
step is taken as $1/100$ of the rf-period. For all cases studied in the following, the radio-frequency  trapping
field is kept constant with a frequency $\Omega_{rf}/2\pi = 10$~MHz, which results in a time step of
$10^{-9}$~s, and an rf amplitude $V_0 = 960$~V in a linear trap of inner radius $r_0 = 2.5$~mm.  The trapped ion
species is  the most abundant calcium isotope ($m = 40$~amu), and for this species, the rf trapping parameters
lead  to a Mathieu parameter $q = 0.187$, which obeys the validity criteria of the adiabatic approximation. The
aspect ratio $\alpha$ of the potential is changed by tuning $\omega_z$ which takes values between $2\pi \times
3.5$~kHz and $2\pi \times 1.15$~MHz.
For Doppler laser cooling, we use the strong dipole transition of calcium ion at 397~nm, which has a natural
linewidth of $\Gamma_0 = 1.43 \times 10^8\ $s$^{-1}$.  To optimize  laser cooling \cite{lett89,metcalfbook}, the
laser detuning is set at half the natural linewidth $\delta_{l} =- \Gamma_0 / 2$.  

The choice for the Rabi frequency results from a compromise between reaching the lowest temperature, which requires $\Omega_r \ll \Gamma_0$ \cite{lett89,metcalfbook} and keeping a high enough photon scattering rate to counteract the heating  induced by the trapping potential, called rf-heating  \cite{blumel89,berkeland98,ryjkov05,zhang07}. This heating is induced by the transfer of energy from the rf power source to the macro-motion kinetic energy because of non-linear resonances \cite{alheit95,vedel98}. These resonances can be induced by higher order contributions in the rf trapping potential (not included in the simulations) and by the Coulomb repulsion between ions \cite{vedel84}. With our trapping configuration, the cooling of the simulated trajectories requires a Rabi frequency of the order of $\Gamma_0$ to counteract rf-heating.

 As the rf-heating rate depends on the
temperature of the sample \cite{ryjkov05,zhang07}, it is possible to reach lower equilibrium temperatures by
varying the Rabi frequency during the cooling process. Indeed, one could imagine starting the experiment with a high Rabi frequency to ensure a high enough scattering rate to reach a temperature for which the rf-heating is negligible \cite{ryjkov05}. Then, in a second step, the Rabi frequency could be decreased to reach a lower equilibrium temperature.  In the present work, we use the same constant Rabi
frequency $\Omega_r=\Gamma_0$ throughout our study in order to focus on the differences in the Coulomb coupling induced by the
anisotropy of the cloud.
The limit temperatures reached by the cooling process control the amplitude of the oscillation  of the ions
around their equilibrium position (the non-driven one). This has an effect on the Coulomb
coupling through the relative positions of the ions. Slightly different results are then expected with another Rabi frequency, but the general behaviour of the coupling efficiency should not fundamentally change.

Initial conditions are built in an identical way for the rf case and the pseudo-potential case associated to the same
rf potential. The aim is to prepare a set of ions in an equilibrium state characterized by a  $1$~K temperature.
For this purpose, a random position and velocity is given to each ion which evolves  under the pseudo-potential
trapping force, the Coulomb forces and the photon absorption/emission momentum kicks. The cooling laser is
already applied during the preparation of the initial state  to avoid a shift in position induced by the
radiation pressure at the beginning of the simulations. To prepare an initial equilibrium state at $1$~K, a
Gaussian thermostat \cite{nose84,hoover82,hoover05} is used, which consists in scaling  each velocity at the
end of a fixed period. Starting from these initial conditions, the cloud is then ready to evolve under the Molecular Dynamics simulations which take place in the rf electric field or its associated pseudo-potential. The conversion of the pseudo-potential into its rf counterpart is associated to a rapid rise of the temperature.
This discontinuity does not
affect the long term equilibrium of the dynamics \cite{schiffer03}.   The time evolution starts by a cooling phase, which lasts less
than 10~ms and takes the system from the initial 1~K to temperatures of the order of few mK. Then the dynamics
becomes stationary and the system can be characterized.  To reach a good precision, each temperature ratio curve
results from the average over at least 15 runs with  different initial conditions. Error bars are evaluated at
$1\sigma$ deviation from  the mean value.

\section{\label{sec:PsCase}In the pseudo-potential approximation}
The structural phase transitions between the  minimum energy configurations calculated for different $\alpha$
have been identified by J.P. Schiffer in  \cite{schiffer93}. The Molecular Dynamics simulations used in that work
do not imply any cooling but a thermalization of the sample by a thermostat. Our Doppler laser cooling
simulations do not allow to reach these low temperatures and are limited by the Doppler cooling  limit (0.5~mK
in the case of calcium ions). Therefore, we observe large thermal fluctuations compared to the configurations
demonstrated in \cite{schiffer93}. Nevertheless, we  refer to J.P. Schiffer's work  to look for correlations
between the geometric configuration of the set of ions and the efficiency of the Coulomb coupling to cool the
degrees of freedom that are not directly laser cooled.

As an illustration  of the geometrical evolution of these configurations, let us consider  a 70 ion system at
very low temperature. For $\alpha < 1.5\times 10^{-3}$, the minimum energy configuration is a  linear structure
which becomes a planar zig-zag for higher $\alpha$ values \cite{morigi04},  until $\alpha \simeq 3.1\times 10^{-3}$. Increasing
$\alpha$ beyond this value results in a twisting of the zig-zag, which forms then a 3D structure. It has
been shown  \cite{schiffer93} that the twisting angle increases with increasing $\alpha$. For  large enough $\alpha$ ($\alpha
\simeq 10$), the configuration is again a 2D structure, this time in the plane orthogonal to the symmetry axis
of the trap. In our simulations, we observe some differences in the equilibrium configurations due to thermal
fluctuations.  For example, line or zig-zag configurations can only be observed by time-averaging of ion
positions. Besides these fluctuation effects we find the same structural transitions between different configurations
as observed in \cite{schiffer93}.

\subsection{\label{sec:PsCase:Vdc=0V} In a  potential with  cylindrical symmetry }
\begin{figure}
    \includegraphics[height=4.4cm]{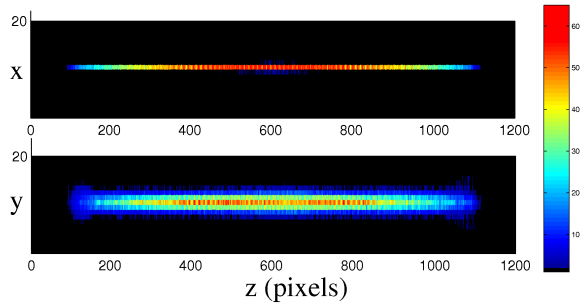}
    \caption{(Color online) Simulated CCD-image  of 70 ions laser cooled along the $x$-axis and trapped in the
    pseudo-potential associated to the rf field defined in \ref{sec:Numerical},  with an anisotropy characterized by
    $\alpha = 10^{-3}$. The pixel size is 1~$\mu$m and the signal associated to one pixel is proportional to the
    number of scattered photons in this pixel, integrated over 10~ms. The picture is projected on the $(x,z)$ plane
    (top) and on the $(y,z)$ plane (bottom).}
    \label{graph_CCDpsa0alpha1e-3}
\end{figure}

In this part, we study the cooling of the three degrees of freedom of ions trapped in the  pseudo-potential
without a static voltage applied on the quadrupole rods ($a_u=0$) and with a single laser beam propagating along
$-x$.  The values of $\alpha$ for which the structural transitions occur   are in good agreement with
those given in \cite{schiffer93}. In the cases where the ions form a line along the $z$ axis, there is a large
amplitude of motion in the plane orthogonal to the laser beam, showing that laser cooling is not efficient in
this plane. A simulated CCD-image of a 70-ion system is shown on Fig.~\ref{graph_CCDpsa0alpha1e-3} resulting from
the projection on  two different planes: the top view shows a picture projected on the $(x,z)$ plane which
includes the laser propagation axis, the bottom view is a projection on the $(y,z)$ plane, orthogonal to the
laser axis. These simulated images show that the thermal fluctuations can be observed  experimentally 
in the direction perpendicular to the trap axis and to the laser beam. A precise analysis of the data reveals that
the amplitude of motion in the directions perpendicular to the laser beam is not only due to thermal motion around the COM but  also to the
motion of the COM itself. The non-cooling of the COM motion in the plane orthogonal to the laser beam has already been analyzed in \cite{hegerfeldt90}. Indeed,  the COM motion is decoupled from the internal Coulomb forces of the cloud. Hence, the COM behaves like a single particle inside the trap potential.
Its kinetic energy is only reduced on the eigen-axis of the trap on which the laser beam has a projection. In the cylindrically symmetric pseudo-potential, the radial directions are all equivalent and the kinetic energy of the COM is reduced only along the laser beam direction.

\begin{figure}
   \includegraphics[width=8cm]{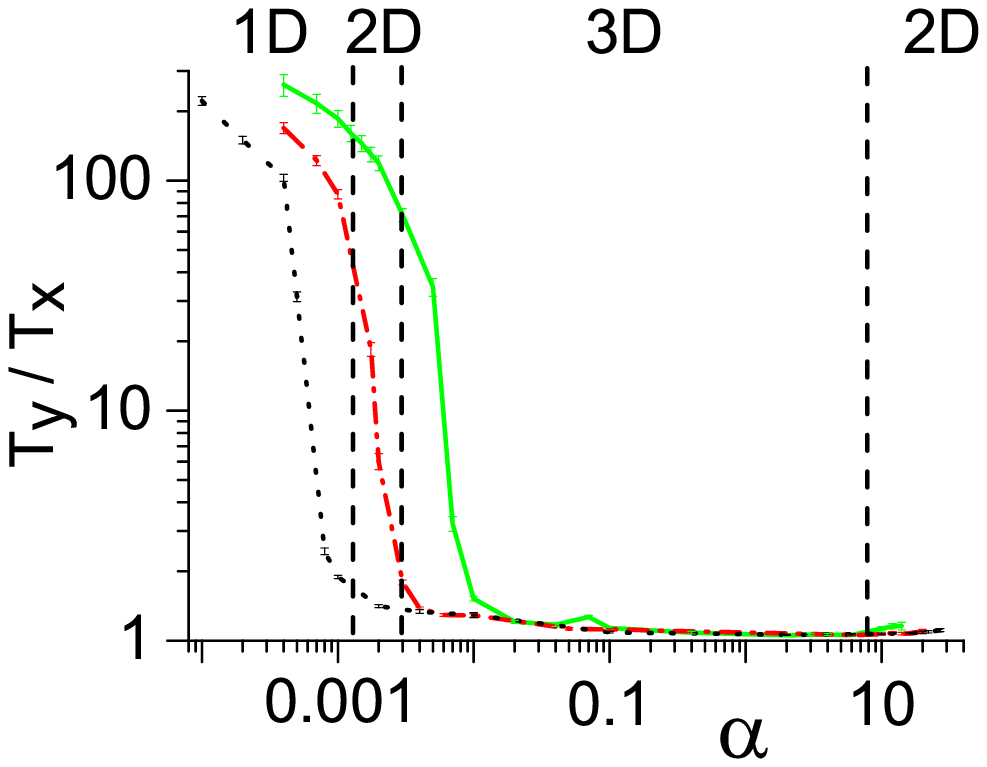} \\
   \includegraphics[width=8cm]{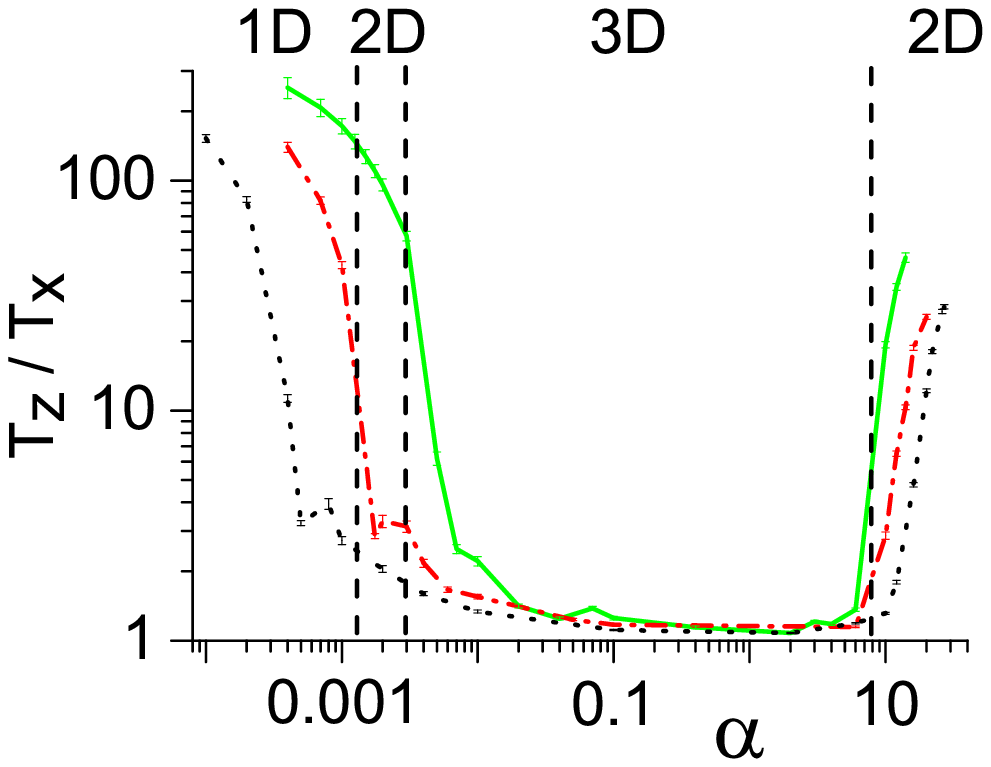}
   \caption{(Color online) Ratio of the temperature in $y$ (top) or $z$ (bottom) direction with respect to the temperature in
   the laser cooled direction, versus the anisotropy of the trap $\alpha$. The trapping potential is the static
   pseudo-potential with cylindrical symmetry ($a_u = 0$). Calculations are made for 35 ions (solid line), 70 ions
   (dash-dotted line) and 140 ions (dotted line). The vertical lines mark the transition between different geometric configurations for 70 ions. }
   \label{graph_ps0alpha}
\end{figure}

The dependance of the temperature ratio reached by laser cooling with the anisotropy of the trapping potential
is summarised in Fig.~\ref{graph_ps0alpha}. For small values of $\alpha$, corresponding to a string of ions, the Coulomb coupling does not lead to  cooling in the $y$ and $z$ direction and the
temperature ratio shown on  Fig.~\ref{graph_ps0alpha} can be as high as 200. The temperature $T_x$ itself
fluctuates around the Doppler cooling limit (0.5~mK).

Increasing $\alpha$ leads to the zig-zag structure identified for very low temperatures, but in our case, the
amplitude of the thermal motion is comparable to the distance between the two arms of the zig-zag. Moreover,
there is no  configuration in the $(x,y)$ plane with minimal energy and the calculated structure undergoes
stochastic rotations around the $z$-axis, which prevent from obtaining an averaged structure comparable to
the one shown on Fig.~\ref{graph_CCDpsa0alpha1e-3}. The extension of the ion configuration in 3D results in a better
coupling between the 3 degrees of freedom which is visible on Fig.~\ref{graph_ps0alpha} with temperature ratios
close to 1. One can notice that the temperature ratio decreases drastically by two orders of magnitude for
$\alpha$ values in the range for which the stable configurations at low temperature switch from a line to a
twisted zig-zag. When $\alpha$ reaches values for which the ions settle in a plane configuration  orthogonal to
the trap axis ($\alpha \ge 10 $ for 70 ions), the spatial extension of the cloud in the $z$ direction is too
small to keep an efficient coupling between the motion along $z$ and the motion in the $(x,y)$ plane. This
translates into a rapid growth of the $T_z/T_x$ ratio whereas the $T_y/T_x$ ratio remains very close to 1 (see
Fig.~\ref{graph_CCDpsa0alpha1e-3}). The stationary regime is reached when the amplitude  of the thermal motion
along $z$ is large enough to recover a strong enough coupling with the laser cooled direction of motion. The
correlations between the 3D geometric extension of the cloud with the temperature ratio  demonstrate the limits
of Coulomb coupling for  efficient cooling in 3D with a single laser beam. 

\subsection{\label{sec:PsCase:Vdc=10V} Without  cylindrical symmetry}
\begin{figure}
    \includegraphics[height = 3cm]{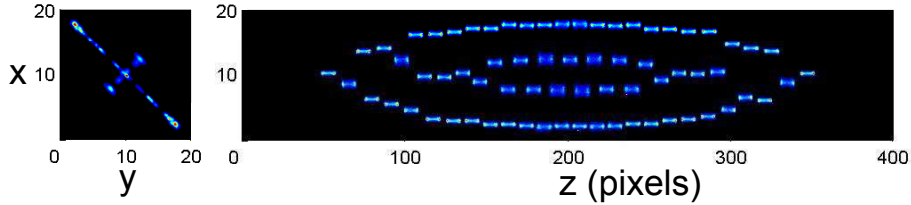}
    \caption{(Color online) Simulated picture  of 70 ions, laser cooled along the $x$-axis and trapped in a
    pseudo-potential with an anisotropy characterized by $\alpha =4 \times 10^{-2}$ and a non symmetric radial
    potential with $a_u=7.805 \times 10^{-3}$. The pixel size is 0.05~$\mu$m in the $x$ and $y$ direction and
    0.2~$\mu$m in the $z$ direction.  The signal associated to one pixel is proportional to the number of
    scattered photons in this pixel, accumulated over 2~ms. The picture is projected on the $(x,y)$ plane (left)
    and on the $(y,z)$ plane (right) and the scale is in $\mu$m.}
    \label{CCD_ps_a10}
\end{figure}

In this part, we break the cylindrical symmetry of the trapping pseudo-potential by a very small amount ($a_u = 7.805 \cdot 10^{-3}$ or $V_{dc}=20$~V) to analyze the effect of the potential asymmetry on the cooling efficiency without notably modifying   the potential depths. As before, we choose a set of 70 ions as a representative case for the descriptions of the dynamics.  For $\alpha \le 1.5 \times 10^{-3}$,  the structure still
forms a line along the $z$ direction. But for larger $\alpha$ values, the dynamics in the radial plane is very different from the  dynamics in a cylindrical symmetric potential. Once cooled, the ions tend to align in the direction of
minimum steepness of the potential, noted $v$ for a positive $V_{dc}$.  For $\alpha = 2\times 10^{-3}$, the structure expands in the $(z,v)$ plane, forming a stationary zig-zag structure. Contrary to the symmetrical case, the structure remains in the same plane and is not subject to chaotic  rotations around the $z$-axis. Increasing $\alpha$ to $2 \times 10^{-2}$, a new 2D structure appears. It results from the already observed zig-zag geometry plus a central line along the $z$-axis. Increasing $\alpha$ further to $4 \times 10^{-2}$ leads to the transition of this central line to another zig-zag configuration in the $ (z,u)$ plane, which is orthogonal to the outer one. A picture of such a system  is shown on  Fig.~\ref{CCD_ps_a10} where the different planes occupied by the two zig-zags are clearly visible. On this picture, the spot associated to each ion is elongated in the $z$ direction due to the COM motion.  Intermediate values of $\alpha$ produce 3D structures without notable symmetries. For large $\alpha$ values ($\alpha \ge 30$) corresponding to a planar structure in the symmetric case, the ions organize themselves around the $v$ axis in the $(u,v)$ plane. In the limit of very large $\alpha$ ($\simeq 800$), the structure tends to be a perfect line along the $v$ axis.

\begin{figure}
    \includegraphics[width=8cm]{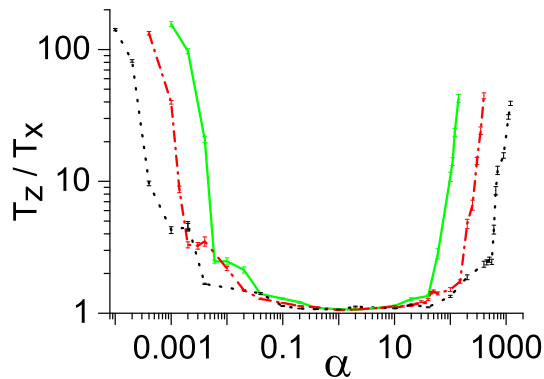} \\
    \caption{(Color online) Ratio of the temperature along the trap axis over the temperature in the laser
    cooled direction, versus the anisotropy of the trap $\alpha$. The trapping potential is the pseudo-potential
    with a broken cylindrical symmetry characterized by $a_u = 7.805 \times 10^{-3}$. Calculations are made for
    35 ions (solid line), 70 ions (dash-dotted line) and 140 ions (dotted line).}
    \label{graph_ps10alpha}
\end{figure}

For any $\alpha$ and any tested number of ions, the temperature ratio $T_y / T_x$ is approximately equal to 1
and does not undergo significant changes. This is attributed to the fact that the laser beam does not propagate
along one of the eigen-axis of the pseudo-potential (see Eq.~\ref{eq_pseudopot}) leading to a strong coupling
between the motion along $x$ and $y$. The thermal fluctuations in the radial plane are then smaller than for the
symmetric case.  For the same reasons, only the  motion of the COM in the $z$ direction is not cooled. The $T_z/
T_x$ ratio given in Fig~\ref{graph_ps10alpha} shows  the same dependance with the anisotropy of the potential as
in the case of a cylindrically symmetric potential. For large $\alpha$ values, the motion in the $z$ direction
decouples from the other motions and the $T_z/ T_x$ ratio increases.

In the particular case of small $\alpha$ values, for which the stable configuration is a line along $z$, we
observe that laser cooling induces a rotation of the ions around the $z$ axis. The combined effect of the
asymmetric  pseudo-potential and of the laser cooling gives birth to a rotation with an orientation depending on
the polarity of the static voltage $V_{dc}$ and the axis of propagation of the laser beam, its direction
being irrelevant. For a positive voltage $V_{dc}$, the maximal rotation is found when the laser is along the
$x$-axis and  it decreases to zero when the laser is aligned with a pair of rods of the trap. When the trap is further rotated compared to the axis of propagation of the laser,
the rotation of the ions increases again but in the opposite direction.  For two counter-propagating laser beams, the calculated
rotation is the same if we model their mechanical effect by momentum kicks, as explained before, or if we
approximate this effect by a friction force \cite{lett89,metcalfbook}. When using two orthogonal laser beams,
no rotation is observed.

\begin{figure}
    \includegraphics[width=7cm]{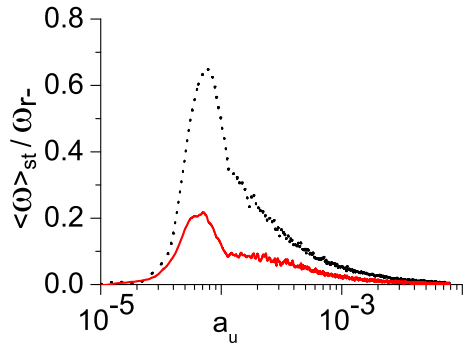}
    \caption{(Color online) Ratio of the stationary mean rotation $\left<\omega \right>$ over $\omega_{r-}$ of a
    system of 70 ions, as a function of the Mathieu parameter $a_u$, for an angle $\phi_{rod} = \pi /4$ and
    $\alpha=10^{-3}$ (dashed line) and $\alpha=2 \times 10^{-3}$ (solid line).
    \label{gr-om_a_ps}}
\end{figure}

The mean rotational frequency $\left<\omega \right>$, calculated  in the  COM frame, depends on the Mathieu parameter $a_u$.
This is illustrated on Fig.~\ref{gr-om_a_ps} for two different values of $\alpha$. For the chosen laser
parameters, the two curves show a maximum  for $a_u \simeq 10^{-4}$  and the rotation becomes negligible for $a_u \ge 0.01$.

The simulations presented in this section demonstrate the importance of the geometric configurations on the
efficiency of the Coulomb coupling to thermalize the degrees of motion. In a perfect symmetric pseudo-potential,
 a single laser beam in the radial plane is not sufficient to cool the three degrees
of freedom of a  cloud of ions when its configuration does not extend in the three directions of space.  Breaking
the cylindrical symmetry of the potential makes  a single laser beam  sufficient to cool the motion in the
radial plane to the Doppler limit. Nevertheless the Doppler limit may not be reached in experiments because of
the rf driven motion which can lead to a heating of the sample \cite{blumel89,prestage91,hornekaer02,ryjkov05}.
This competition between cooling and heating can modify the efficiency of the Coulomb  interaction to thermalize
different degrees of freedom. Therefore, in the next section,  the results  obtained in the
pseudo-potential approximation are compared to those obtained with the full rf electric field.

\section{\label{sec:RfCase}In the radio-frequency electric field}

The ideal potential used in the simulations can not be reproduced in experiments  because of mechanical
imperfections of the trap, induced capacity, rf phase mismatch   or material deposition on the electrodes \cite{berkeland98}. Hence, dynamics related  
in the following for a pure radio-frequency field may not be observed in the laboratory. In the second part of this section, an additional  static voltage  takes into account  asymmetries of the potential induced by  imperfections.

\subsection{In a pure rf electric field \label{sec:pureRf}}
\begin{figure}
    \includegraphics[height=7cm]{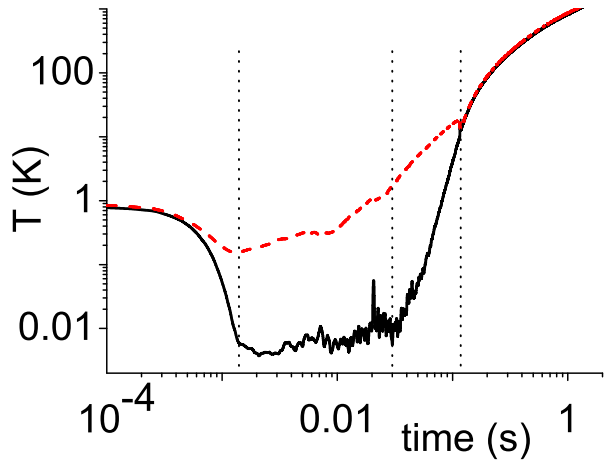}
    \caption{(Color online) Time evolution of the temperature in the $x$ (solid line) and $y$ (dashed line) direction of motion of 70 laser cooled ions in a linear rf potential with anisotropy parameter $\alpha = 7\times 10^{-4}$. The vertical lines mark the four evolution phases mentioned in the text.  }
    \label{graph_temp_rf0}
\end{figure}

The Molecular Dynamics simulations of laser-cooled ions in a pure rf quadrupole field can be split into four
successive temporal phases. First, the ion cloud undergoes  a cooling phase until it forms a dense cloud, with a
shape depending on the anisotropy of the trap. The  temperatures $T_x$ and $T_y$ typically reach   values
between 1 and 10~mK and an example of the time evolution of these temperatures is shown in
Fig.~\ref{graph_temp_rf0}.   For small enough $\alpha$  ($\alpha \leq 10^{-3}$), for which a linear configuration
along the $z$-axis is expected from \cite{schiffer93}, the amplitude of motion along $y$ is due to thermal
motion and to the COM motion, as in the pseudo-potential case. But contrary to the previous case, the plane
occupied by the ions during their motion is no more static. It now forms an angle with the $y$-axis oscillating
at the frequency of the rf electric field. The introduction of a  second counter-propagating laser beam
suppresses these oscillations without modifying the other features described in the following.

In a second step, after being reduced by laser cooling, the amplitude of motion increases. This  process is not
observed in the pseudo-potential case and is attributed to rf heating~\cite{drewsen98}. In the mean time,
the COM motion increases in the $y$ direction and no geometric configuration can be identified.

\begin{figure}
    \includegraphics[height=7cm]{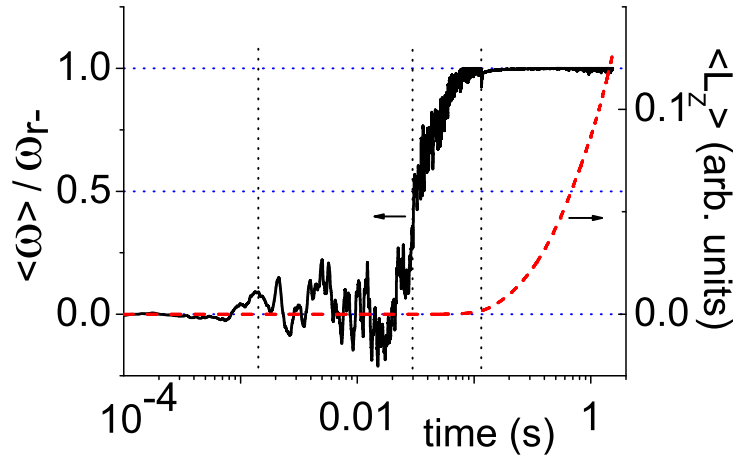}
    \caption{(Color online) Time evolution of the ratio of the mean rotation $\left<\omega \right>$ over
    $\omega_{r}$ (left axis) and angular momentum $\left<\overline{L}_z\right>$  (right axis, arbitrary units),
    calculated in the COM frame, for a sample of 70 ions  in a linear rf potential with anisotropy parameter
    $\alpha = 7\times 10^{-4}$. The vertical lines mark the four evolution phases mentioned in the text.}
    \label{graph_om_rf0}
\end{figure}

The third phase of the time evolution starts typically 30~ms after the beginning of the experiment with the
amplification of the COM motion in the laser direction. In the ion motion in the
radial plane, one can identify the rotation of the COM   around the trap center,  superposed to the rotation of the whole cloud around its COM
(see Fig.~\ref{graph_om_rf0}).  The sign of these angular frequencies is random. At the end of this third phase, the contributions of the kinetic energies
associated to these two motions  are of the same order.

The onset of the last part of the dynamics, taking place 0.1 s after the beginning of the experiment, is
controlled by the mean rotation of the cloud   in the COM frame $\left<\omega \right>$. At that time, its angular frequency
 exactly matches the frequency of the pseudo-potential  $\omega_r$ and  remains constant.  The cloud
then expands in the radial direction, as illustrated on Fig.~\ref{graph_om_rf0} where the mean angular momentum
in the COM frame,  $\left<L_z\right>$ is plotted with $\left<\omega \right>$.

\subsection{With an added static asymmetric potential \label{sec:RfStatic}}
\begin{figure}
   \includegraphics[width=8cm]{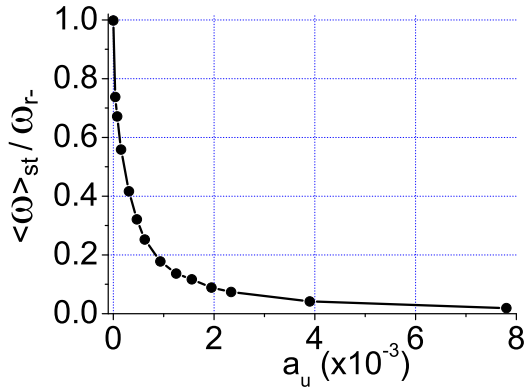}
    \caption{Ratio of the stationary mean rotation $\left<\omega \right>_{st}$ over $\omega_{r-}$ of a 70 ions
    cloud in the COM frame,  versus the Mathieu parameter $a_u$,  in a linear rf potential with anisotropy
    parameter $\alpha = 7\times 10^{-4}$.}
    \label{graph_omega_a}
\end{figure}

We assume this particular dynamics is made possible by the perfect symmetry of the rf field and that breaking
the symmetry could change the dynamics completely.  To test this assumption, we  add a static potential $V_{dc}$
to the rf electrodes, as for the pseudo-potential case developed in \ref{sec:PsCase:Vdc=10V}. When increasing
$V_{dc}$, we observe an important decrease of the stationary mean   rotation $\left<\omega \right>_{st}$,
reported on Fig.~\ref{graph_omega_a}. The sign of the rotation   is now fixed by the polarity  of the static
voltage and the orientation  of the laser beam, which is clockwise for positive $V_{dc}$ and a laser beam along
the $x$ axis, which is consistent with the observations made in \ref{sec:PsCase:Vdc=10V}.

Figure~\ref{graph_omega_a} shows that for a Mathieu parameter  $a_u = 7.805\times 10^{-3}$, corresponding to
$V_{dc}=20$~V, the rotation can be neglected. The dynamics  reaching an equilibrium, it is then possible to
calculate the temperature for the three components of motion. 

 As for laser cooling, the anisotropy of the trapping potential influences the rf
heating rate through the spatial distribution of the ions inside the trap. Therefore, depending on the
anisotropy $\alpha$, the competition between cooling and heating processes may not always lead to a cooling of
the sample. The cooling rate can be tuned, to a certain extent, by the atom-laser coupling parameters introduced
in \ref{sec:dynamics}. Nevertheless, to focus on the effect of the geometric configurations,  we keep these
parameters as defined in \ref{sec:Numerical} throughout all the simulations presented here. In practice, this
leads to the systematic  cooling of the sample only for $\alpha$ smaller than 0.4.  For larger $\alpha$, the net cooling efficiency depends
on the initial conditions and increasing $\alpha$ lowers the probability to reach a cold sample, demonstrating
that rf heating increases with $\alpha$.

To pursue the study for larger alpha values, we start with a cold cloud obtained for $\alpha = 0.4$ and  increase $\alpha$ slowly enough to make sure that  the equilibrium is reached when the temperatures are measured.

\begin{figure}
    \includegraphics[width=8cm]{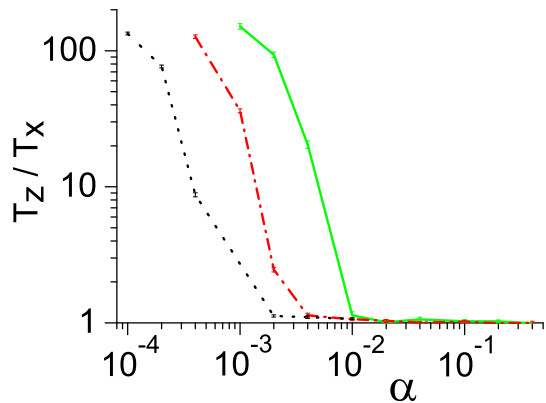}
    \caption{(Color online) Ratio of the temperature in the uncooled direction $T_z$ over the temperature in the laser cooled
    direction $T_x$, versus the anisotropy of the trap $\alpha$. The trapping potential is the rf potential plus
    a dc contribution characterized by $a_u = 7.805 \cdot 10^{-3}$. Calculations are made for 35 ions (solid
    line), 70 ions (dash-dotted line) and 140 ions (dotted line).}
    \label{graph_rfalpha}
\end{figure}

As for the pseudo-potential case for $a_u \neq 0$,
the motions along $x$ and $y$ show the same temperature, but contrary to the former case, this temperature
does not reach the Doppler limit and varies between 1 and 100~mK, depending on $\alpha$ and the number of ions.
The temperature ratio $T_z/T_x$ is plotted on Fig.~\ref{graph_rfalpha} for $\alpha< 0.4$ and for the particular case $a_u = 7.805\
10^{-3}$ and shows the same behaviour as for the pseudo-potential case of Fig.~\ref{graph_ps10alpha}. Indeed,
the structural configurations are very similar in both cases. For $0.4 < \alpha < 200$ and a 70 ion cloud, the temperature ratio $T_z/T_x$ is found to be equal to one. Temperature values increase from 10 to 100~mK when $\alpha$ is increased from 1 to 2. They abruptly fall to 3~mK when $\alpha$ reaches 35. This corresponds  to a transition to a structure organized along the $v$ axis.  When $\alpha$ reaches 200, we observe the same rise of $T_z$ as in the pseudo-potential case. A close look at the ion configuration reveals that this corresponds to the transition to a string along the $v$ axis. 

The results of these numerical studies are twofold. First, they show that in an ideal linear rf trap, ions that
are laser cooled along one radial direction can not be confined because of a rotation of the cloud leading to
its expansion.  Second, they show that a very small asymmetry in the radial trapping field is sufficient to inhibit
this rotation and keep the ions in a stationary structure. In this case, the temperature ratio shows the same
dependence with the anisotropy $\alpha$  as has been observed in the pseudo-potential.

\section{Conclusion}
In this article, we  study by means of Molecular Dynamics simulations, the dynamics of ions in a linear
quadrupolar trap,  cooled by a single laser beam propagating in the radial plane. The dependence of the
thermalization of the three degrees of motion with the spatial configuration of the ions is analyzed. These
simulations  show  the determinant role of Coulomb coupling in the thermalization process of the degrees of
freedom orthogonal to the laser beam.  When the trapping
potential is modeled by its static pseudo-potential, we observe a strong correlation between the spatial
configuration of the ions and the thermalization efficiency. Actually, the temperatures in the three directions
of motion reach the same limit only if the cloud has a 3D extension. When taking into account the full rf
electric field,  a rotation of the cloud can be observed, that leads to its expansion and thus prevents any calculations
of stationary properties. This rotation can be suppressed by breaking the symmetry of the trapping potential,
which is realized by applying a static voltage to the rf electrodes. This asymmetry  efficiently couples
the two radial degrees of freedom which then reach the same limit temperature. When the simulations lead to cold
samples, the calculated geometric structures and temperature ratios  are equivalent in the pseudo-potential  and
in the full rf potential. In this last case, for the chosen laser cooling parameters and starting with a 1~K cloud, rf heating
prevents to reach a cold sample when the anisotropy parameter $\alpha$ leads to a large extension of the cloud
in the radial plane. Nevertheless, obtaining  cold samples  with high $\alpha$ values (up to 300) was made possible by increasing $\alpha$ once the sample cooled. The   asymmetry  required to prevent the rotation of the cloud is small enough to be
encountered in any realistic experiment, which explains the possibility of stable trapping under single beam laser-cooling.

\begin{acknowledgments}
M.M. gratefully thanks Yves Elskens for very helpful and stimulating discussions. The authors would like to
acknowledge fruitful discussions with Fernande Vedel and Florent Calvo.
\end{acknowledgments}


\begin{thebibliography}{42}
\expandafter\ifx\csname natexlab\endcsname\relax\def\natexlab#1{#1}\fi
\expandafter\ifx\csname bibnamefont\endcsname\relax
  \def\bibnamefont#1{#1}\fi
\expandafter\ifx\csname bibfnamefont\endcsname\relax
  \def\bibfnamefont#1{#1}\fi
\expandafter\ifx\csname citenamefont\endcsname\relax
  \def\citenamefont#1{#1}\fi
\expandafter\ifx\csname url\endcsname\relax
  \def\url#1{\texttt{#1}}\fi
\expandafter\ifx\csname urlprefix\endcsname\relax\def\urlprefix{URL }\fi
\providecommand{\bibinfo}[2]{#2}
\providecommand{\eprint}[2][]{\url{#2}}

\bibitem[{\citenamefont{H{\"a}nsch and Schawlow}(1975)}]{hansch75}
\bibinfo{author}{\bibfnamefont{T.}~\bibnamefont{H{\"a}nsch}} \bibnamefont{and}
  \bibinfo{author}{\bibfnamefont{A.}~\bibnamefont{Schawlow}},
  \bibinfo{journal}{Optics Comm.} \textbf{\bibinfo{volume}{13}},
  \bibinfo{pages}{68} (\bibinfo{year}{1975}).

\bibitem[{\citenamefont{Wineland and Dehmelt}(1975)}]{wineland75}
\bibinfo{author}{\bibfnamefont{D.~J.} \bibnamefont{Wineland}} \bibnamefont{and}
  \bibinfo{author}{\bibfnamefont{H.~G.} \bibnamefont{Dehmelt}},
  \bibinfo{journal}{Bull. Am. Phys. Soc.} \textbf{\bibinfo{volume}{20}},
  \bibinfo{pages}{637} (\bibinfo{year}{1975}).

\bibitem[{\citenamefont{Paul}(1990)}]{paul90}
\bibinfo{author}{\bibfnamefont{W.}~\bibnamefont{Paul}}, \bibinfo{journal}{Rev.
  Mod. Phys.} \textbf{\bibinfo{volume}{62}}, \bibinfo{pages}{531}
  (\bibinfo{year}{1990}).

\bibitem[{\citenamefont{Neuhauser et~al.}(1980)\citenamefont{Neuhauser,
  Hohenstatt, Toschek, and Dehmelt}}]{neuhauser80}
\bibinfo{author}{\bibfnamefont{W.}~\bibnamefont{Neuhauser}},
  \bibinfo{author}{\bibfnamefont{M.}~\bibnamefont{Hohenstatt}},
  \bibinfo{author}{\bibfnamefont{P.}~\bibnamefont{Toschek}}, \bibnamefont{and}
  \bibinfo{author}{\bibfnamefont{H.}~\bibnamefont{Dehmelt}},
  \bibinfo{journal}{Phys. Rev. A} \textbf{\bibinfo{volume}{22}},
  \bibinfo{pages}{1137} (\bibinfo{year}{1980}).

\bibitem[{\citenamefont{Wineland et~al.}(1987)\citenamefont{Wineland, Itano,
  Bergquist, and Hulet}}]{wineland87}
\bibinfo{author}{\bibfnamefont{D.~J.} \bibnamefont{Wineland}},
  \bibinfo{author}{\bibfnamefont{W.~M.} \bibnamefont{Itano}},
  \bibinfo{author}{\bibfnamefont{J.~C.} \bibnamefont{Bergquist}},
  \bibnamefont{and} \bibinfo{author}{\bibfnamefont{R.~G.} \bibnamefont{Hulet}},
  \bibinfo{journal}{Phys. Rev. A} \textbf{\bibinfo{volume}{36}},
  \bibinfo{pages}{2220} (\bibinfo{year}{1987}).

\bibitem[{\citenamefont{Bergquist et~al.}(1987)\citenamefont{Bergquist, Itano,
  and Wineland}}]{bergquist87}
\bibinfo{author}{\bibfnamefont{J.~C.} \bibnamefont{Bergquist}},
  \bibinfo{author}{\bibfnamefont{W.~M.} \bibnamefont{Itano}}, \bibnamefont{and}
  \bibinfo{author}{\bibfnamefont{D.~J.} \bibnamefont{Wineland}},
  \bibinfo{journal}{Phys. Rev. A} \textbf{\bibinfo{volume}{36}},
  \bibinfo{pages}{428} (\bibinfo{year}{1987}).

\bibitem[{\citenamefont{Diedrich et~al.}(1989)\citenamefont{Diedrich,
  Bergquist, Itano, and Wineland}}]{diedrich89}
\bibinfo{author}{\bibfnamefont{F.}~\bibnamefont{Diedrich}},
  \bibinfo{author}{\bibfnamefont{J.~C.} \bibnamefont{Bergquist}},
  \bibinfo{author}{\bibfnamefont{W.~M.} \bibnamefont{Itano}}, \bibnamefont{and}
  \bibinfo{author}{\bibfnamefont{D.~J.} \bibnamefont{Wineland}},
  \bibinfo{journal}{Phys. Rev. Lett.} \textbf{\bibinfo{volume}{62}},
  \bibinfo{pages}{403} (\bibinfo{year}{1989}).

\bibitem[{\citenamefont{Monroe et~al.}(1995)\citenamefont{Monroe, Meekhof,
  King, Itano, and Wineland}}]{monroe95}
\bibinfo{author}{\bibfnamefont{C.}~\bibnamefont{Monroe}},
  \bibinfo{author}{\bibfnamefont{D.~M.} \bibnamefont{Meekhof}},
  \bibinfo{author}{\bibfnamefont{B.~E.} \bibnamefont{King}},
  \bibinfo{author}{\bibfnamefont{W.~M.} \bibnamefont{Itano}}, \bibnamefont{and}
  \bibinfo{author}{\bibfnamefont{D.~J.} \bibnamefont{Wineland}},
  \bibinfo{journal}{Phys. Rev. Lett.} \textbf{\bibinfo{volume}{75}},
  \bibinfo{pages}{4714} (\bibinfo{year}{1995}).

\bibitem[{\citenamefont{Wineland and Itano}(1979)}]{wineland79}
\bibinfo{author}{\bibfnamefont{D.~J.} \bibnamefont{Wineland}} \bibnamefont{and}
  \bibinfo{author}{\bibfnamefont{W.~M.} \bibnamefont{Itano}},
  \bibinfo{journal}{Phys. Rev. A} \textbf{\bibinfo{volume}{20}},
  \bibinfo{pages}{1521} (\bibinfo{year}{1979}).

\bibitem[{\citenamefont{Schiffer}(1993)}]{schiffer93}
\bibinfo{author}{\bibfnamefont{J.~P.} \bibnamefont{Schiffer}},
  \bibinfo{journal}{Phys. Rev. Lett.} \textbf{\bibinfo{volume}{70}},
  \bibinfo{pages}{818} (\bibinfo{year}{1993}).

\bibitem[{\citenamefont{Dubin}(1993)}]{dubin93b}
\bibinfo{author}{\bibfnamefont{D.~H.~E.} \bibnamefont{Dubin}},
  \bibinfo{journal}{Phys. Rev. Lett.} \textbf{\bibinfo{volume}{71}},
  \bibinfo{pages}{2753} (\bibinfo{year}{1993}).

\bibitem[{\citenamefont{Morigi et~al.}(1999)\citenamefont{Morigi, Eschner,
  Cirac, and Zoller}}]{morigi99}
\bibinfo{author}{\bibfnamefont{G.}~\bibnamefont{Morigi}},
  \bibinfo{author}{\bibfnamefont{J.}~\bibnamefont{Eschner}},
  \bibinfo{author}{\bibfnamefont{J.~I.} \bibnamefont{Cirac}}, \bibnamefont{and}
  \bibinfo{author}{\bibfnamefont{P.}~\bibnamefont{Zoller}},
  \bibinfo{journal}{Phys. Rev. A} \textbf{\bibinfo{volume}{59}},
  \bibinfo{pages}{3797} (\bibinfo{year}{1999}).

\bibitem[{\citenamefont{Morigi and Eschner}(2001)}]{morigi01}
\bibinfo{author}{\bibfnamefont{G.}~\bibnamefont{Morigi}} \bibnamefont{and}
  \bibinfo{author}{\bibfnamefont{J.}~\bibnamefont{Eschner}},
  \bibinfo{journal}{Phys. Rev. A} \textbf{\bibinfo{volume}{64}},
  \bibinfo{pages}{063407} (\bibinfo{year}{2001}).

\bibitem[{\citenamefont{Morigi and Eschner}(2003)}]{morigi03b}
\bibinfo{author}{\bibfnamefont{G.}~\bibnamefont{Morigi}} \bibnamefont{and}
  \bibinfo{author}{\bibfnamefont{J.}~\bibnamefont{Eschner}},
  \bibinfo{journal}{J. Phys. B} \textbf{\bibinfo{volume}{36}},
  \bibinfo{pages}{1041} (\bibinfo{year}{2003}).

\bibitem[{\citenamefont{Javanainen}(1986)}]{javanainen86}
\bibinfo{author}{\bibfnamefont{J.}~\bibnamefont{Javanainen}},
  \bibinfo{journal}{Phys. Rev. Lett.} \textbf{\bibinfo{volume}{56}},
  \bibinfo{pages}{1798} (\bibinfo{year}{1986}).

\bibitem[{\citenamefont{DeVoe et~al.}(1989)\citenamefont{DeVoe, Hoffnagle, and
  Brewer}}]{devoe89}
\bibinfo{author}{\bibfnamefont{R.~G.} \bibnamefont{DeVoe}},
  \bibinfo{author}{\bibfnamefont{J.}~\bibnamefont{Hoffnagle}},
  \bibnamefont{and} \bibinfo{author}{\bibfnamefont{R.~G.}
  \bibnamefont{Brewer}}, \bibinfo{journal}{Phys. Rev. A}
  \textbf{\bibinfo{volume}{39}}, \bibinfo{pages}{4362} (\bibinfo{year}{1989}).

\bibitem[{\citenamefont{Cirac et~al.}(1994)\citenamefont{Cirac, Garay, Blatt,
  Parkins, and Zoller}}]{cirac94}
\bibinfo{author}{\bibfnamefont{J.~I.} \bibnamefont{Cirac}},
  \bibinfo{author}{\bibfnamefont{L.~J.} \bibnamefont{Garay}},
  \bibinfo{author}{\bibfnamefont{R.}~\bibnamefont{Blatt}},
  \bibinfo{author}{\bibfnamefont{A.~S.} \bibnamefont{Parkins}},
  \bibnamefont{and} \bibinfo{author}{\bibfnamefont{P.}~\bibnamefont{Zoller}},
  \bibinfo{journal}{Phys. Rev. A} \textbf{\bibinfo{volume}{49}},
  \bibinfo{pages}{421} (\bibinfo{year}{1994}).

\bibitem[{\citenamefont{Pedregosa et~al.}(2010)\citenamefont{Pedregosa,
  C.Champenois, Houssin, and M.Knoop}}]{pedregosa10}
\bibinfo{author}{\bibfnamefont{J.}~\bibnamefont{Pedregosa}},
  \bibinfo{author}{\bibnamefont{C.Champenois}},
  \bibinfo{author}{\bibfnamefont{M.}~\bibnamefont{Houssin}}, \bibnamefont{and}
  \bibinfo{author}{\bibnamefont{M.Knoop}}, \bibinfo{journal}{Int. J. Mass
  Spec.} \textbf{\bibinfo{volume}{290}}, \bibinfo{pages}{100}
  (\bibinfo{year}{2010}).

\bibitem[{\citenamefont{McLachlan}(1947)}]{McLachlan47}
\bibinfo{author}{\bibfnamefont{N.}~\bibnamefont{McLachlan}},
  \emph{\bibinfo{title}{Theory and Application of Mathieu Functions}}
  (\bibinfo{publisher}{Clarendon, Oxford}, \bibinfo{year}{1947}).

\bibitem[{\citenamefont{Ghosh}(1995)}]{gosh_book}
\bibinfo{author}{\bibfnamefont{P.~K.} \bibnamefont{Ghosh}},
  \emph{\bibinfo{title}{Ion traps}} (\bibinfo{publisher}{Oxford university
  press}, \bibinfo{year}{1995}).

\bibitem[{\citenamefont{Drewsen and Br{\o}ner}(2000)}]{drewsen00}
\bibinfo{author}{\bibfnamefont{M.}~\bibnamefont{Drewsen}} \bibnamefont{and}
  \bibinfo{author}{\bibfnamefont{A.}~\bibnamefont{Br{\o}ner}},
  \bibinfo{journal}{Phys. Rev. A} \textbf{\bibinfo{volume}{62}},
  \bibinfo{pages}{045401} (\bibinfo{year}{2000}).

\bibitem[{\citenamefont{Dehmelt}(1967)}]{dehmelt67}
\bibinfo{author}{\bibfnamefont{H.}~\bibnamefont{Dehmelt}},
  \bibinfo{journal}{Advances in Atomic and Molecular Physics}
  \textbf{\bibinfo{volume}{3}}, \bibinfo{pages}{53} (\bibinfo{year}{1967}).

\bibitem[{\citenamefont{Bl\"umel et~al.}(1988)\citenamefont{Bl\"umel, Chen,
  Peik, Quint, Schleich, Shen, and Walther}}]{blumel88}
\bibinfo{author}{\bibfnamefont{R.}~\bibnamefont{Bl\"umel}},
  \bibinfo{author}{\bibfnamefont{J.}~\bibnamefont{Chen}},
  \bibinfo{author}{\bibfnamefont{E.}~\bibnamefont{Peik}},
  \bibinfo{author}{\bibfnamefont{W.}~\bibnamefont{Quint}},
  \bibinfo{author}{\bibfnamefont{W.}~\bibnamefont{Schleich}},
  \bibinfo{author}{\bibfnamefont{Y.}~\bibnamefont{Shen}}, \bibnamefont{and}
  \bibinfo{author}{\bibfnamefont{H.}~\bibnamefont{Walther}},
  \bibinfo{journal}{Nature} \textbf{\bibinfo{volume}{334}},
  \bibinfo{pages}{309} (\bibinfo{year}{1988}).

\bibitem[{\citenamefont{Schiffer}(2003)}]{schiffer03}
\bibinfo{author}{\bibfnamefont{J.~P.} \bibnamefont{Schiffer}},
  \bibinfo{journal}{J. Phys. B} \textbf{\bibinfo{volume}{36}},
  \bibinfo{pages}{511} (\bibinfo{year}{2003}).

\bibitem[{\citenamefont{Schiller and L{\"a}mmerzahl}(2003)}]{schiller03}
\bibinfo{author}{\bibfnamefont{S.}~\bibnamefont{Schiller}} \bibnamefont{and}
  \bibinfo{author}{\bibfnamefont{C.}~\bibnamefont{L{\"a}mmerzahl}},
  \bibinfo{journal}{Phys. Rev. A} \textbf{\bibinfo{volume}{68}},
  \bibinfo{pages}{053406} (\bibinfo{year}{2003}).

\bibitem[{\citenamefont{Zhang et~al.}(2007)\citenamefont{Zhang, Offenberg,
  Roth, Wilson, and Schiller}}]{zhang07}
\bibinfo{author}{\bibfnamefont{C.~B.} \bibnamefont{Zhang}},
  \bibinfo{author}{\bibfnamefont{D.}~\bibnamefont{Offenberg}},
  \bibinfo{author}{\bibfnamefont{B.}~\bibnamefont{Roth}},
  \bibinfo{author}{\bibfnamefont{M.~A.} \bibnamefont{Wilson}},
  \bibnamefont{and} \bibinfo{author}{\bibfnamefont{S.}~\bibnamefont{Schiller}},
  \bibinfo{journal}{Phys. Rev. A} \textbf{\bibinfo{volume}{76}},
  \bibinfo{eid}{012719} (pages~\bibinfo{numpages}{13}) (\bibinfo{year}{2007}).

\bibitem[{\citenamefont{Lett et~al.}(1989)\citenamefont{Lett, Phillips,
  Rolston, Tanner, Watts, and C.I.Westbrook}}]{lett89}
\bibinfo{author}{\bibfnamefont{P.}~\bibnamefont{Lett}},
  \bibinfo{author}{\bibfnamefont{W.}~\bibnamefont{Phillips}},
  \bibinfo{author}{\bibfnamefont{S.}~\bibnamefont{Rolston}},
  \bibinfo{author}{\bibfnamefont{C.}~\bibnamefont{Tanner}},
  \bibinfo{author}{\bibfnamefont{R.}~\bibnamefont{Watts}}, \bibnamefont{and}
  \bibinfo{author}{\bibnamefont{C.I.Westbrook}}, \bibinfo{journal}{J. Opt. Soc.
  Am. B} \textbf{\bibinfo{volume}{6}}, \bibinfo{pages}{2084}
  (\bibinfo{year}{1989}).

\bibitem[{\citenamefont{Metcalf and van~der Straten}(1999)}]{metcalfbook}
\bibinfo{author}{\bibfnamefont{H.}~\bibnamefont{Metcalf}} \bibnamefont{and}
  \bibinfo{author}{\bibfnamefont{P.}~\bibnamefont{van~der Straten}},
  \emph{\bibinfo{title}{Laser cooling and trapping}}
  (\bibinfo{publisher}{Springer}, \bibinfo{year}{1999}).

\bibitem[{\citenamefont{Bl\"umel et~al.}(1989)\citenamefont{Bl\"umel, Kappler,
  Quint, and Walter}}]{blumel89}
\bibinfo{author}{\bibfnamefont{R.}~\bibnamefont{Bl\"umel}},
  \bibinfo{author}{\bibfnamefont{C.}~\bibnamefont{Kappler}},
  \bibinfo{author}{\bibfnamefont{W.}~\bibnamefont{Quint}}, \bibnamefont{and}
  \bibinfo{author}{\bibfnamefont{H.}~\bibnamefont{Walter}},
  \bibinfo{journal}{Phys. Rev.~A} \textbf{\bibinfo{volume}{40}},
  \bibinfo{pages}{808} (\bibinfo{year}{1989}).

\bibitem[{\citenamefont{Berkeland et~al.}(1998)\citenamefont{Berkeland, Miller,
  Bergquist, Itano, and Wineland}}]{berkeland98}
\bibinfo{author}{\bibfnamefont{D.}~\bibnamefont{Berkeland}},
  \bibinfo{author}{\bibfnamefont{J.}~\bibnamefont{Miller}},
  \bibinfo{author}{\bibfnamefont{J.}~\bibnamefont{Bergquist}},
  \bibinfo{author}{\bibfnamefont{W.}~\bibnamefont{Itano}}, \bibnamefont{and}
  \bibinfo{author}{\bibfnamefont{D.}~\bibnamefont{Wineland}},
  \bibinfo{journal}{J. Appl. Phys.} \textbf{\bibinfo{volume}{83}},
  \bibinfo{pages}{5025} (\bibinfo{year}{1998}).

\bibitem[{\citenamefont{Ryjkov et~al.}(2005)\citenamefont{Ryjkov, Zhao, and
  Schuessler}}]{ryjkov05}
\bibinfo{author}{\bibfnamefont{V.~L.} \bibnamefont{Ryjkov}},
  \bibinfo{author}{\bibfnamefont{X.}~\bibnamefont{Zhao}}, \bibnamefont{and}
  \bibinfo{author}{\bibfnamefont{H.~A.} \bibnamefont{Schuessler}},
  \bibinfo{journal}{Phys. Rev.~A} \textbf{\bibinfo{volume}{71}},
  \bibinfo{eid}{033414} (pages~\bibinfo{numpages}{4}) (\bibinfo{year}{2005}).

\bibitem[{\citenamefont{Alheit et~al.}(1995)\citenamefont{Alheit, Hennig,
  Morgenstern, Vedel, and Werth}}]{alheit95}
\bibinfo{author}{\bibfnamefont{R.}~\bibnamefont{Alheit}},
  \bibinfo{author}{\bibfnamefont{C.}~\bibnamefont{Hennig}},
  \bibinfo{author}{\bibnamefont{Morgenstern}},
  \bibinfo{author}{\bibfnamefont{F.}~\bibnamefont{Vedel}}, \bibnamefont{and}
  \bibinfo{author}{\bibfnamefont{G.}~\bibnamefont{Werth}},
  \bibinfo{journal}{Appl. Phys. B} \textbf{\bibinfo{volume}{61}},
  \bibinfo{pages}{277} (\bibinfo{year}{1995}).

\bibitem[{\citenamefont{Vedel et~al.}(1998)\citenamefont{Vedel, Rocher, Knoop,
  and Vedel}}]{vedel98}
\bibinfo{author}{\bibfnamefont{M.}~\bibnamefont{Vedel}},
  \bibinfo{author}{\bibfnamefont{J.}~\bibnamefont{Rocher}},
  \bibinfo{author}{\bibfnamefont{M.}~\bibnamefont{Knoop}}, \bibnamefont{and}
  \bibinfo{author}{\bibfnamefont{F.}~\bibnamefont{Vedel}},
  \bibinfo{journal}{Appl. Phys. B} \textbf{\bibinfo{volume}{66}},
  \bibinfo{pages}{191} (\bibinfo{year}{1998}).

\bibitem[{\citenamefont{Vedel and Andr\'e}(1984)}]{vedel84}
\bibinfo{author}{\bibfnamefont{F.}~\bibnamefont{Vedel}} \bibnamefont{and}
  \bibinfo{author}{\bibfnamefont{J.}~\bibnamefont{Andr\'e}},
  \bibinfo{journal}{Phys. Rev. A} \textbf{\bibinfo{volume}{29}},
  \bibinfo{pages}{2098} (\bibinfo{year}{1984}).

\bibitem[{\citenamefont{Nos\'e}(1984)}]{nose84}
\bibinfo{author}{\bibfnamefont{S.}~\bibnamefont{Nos\'e}}, \bibinfo{journal}{J.
  Chem. Phys.} \textbf{\bibinfo{volume}{81}}, \bibinfo{pages}{511}
  (\bibinfo{year}{1984}).

\bibitem[{\citenamefont{Hoover et~al.}(1982)\citenamefont{Hoover, Ladd, and
  Moran}}]{hoover82}
\bibinfo{author}{\bibfnamefont{W.~G.} \bibnamefont{Hoover}},
  \bibinfo{author}{\bibfnamefont{A.~J.~C.} \bibnamefont{Ladd}},
  \bibnamefont{and} \bibinfo{author}{\bibfnamefont{B.}~\bibnamefont{Moran}},
  \bibinfo{journal}{Phys. Rev. Lett.} \textbf{\bibinfo{volume}{48}},
  \bibinfo{pages}{1818} (\bibinfo{year}{1982}).

\bibitem[{\citenamefont{Hoover}(2005)}]{hoover05}
\bibinfo{author}{\bibfnamefont{W.~G.} \bibnamefont{Hoover}}, in
  \emph{\bibinfo{booktitle}{Statistical physics and beyond: Proceedings of
  2$^{nd}$ mexican meeting on mathematical and experimental physics}}, edited
  by \bibinfo{editor}{\bibfnamefont{F.}~\bibnamefont{Uribe}},
  \bibinfo{editor}{\bibfnamefont{L.}~\bibnamefont{Garcia-Colin}},
  \bibnamefont{and}
  \bibinfo{editor}{\bibfnamefont{E.}~\bibnamefont{Diaz-Herrera}}
  (\bibinfo{year}{2005}), p.~\bibinfo{pages}{16}.

\bibitem[{\citenamefont{Morigi and Fishman}(2004)}]{morigi04}
\bibinfo{author}{\bibfnamefont{G.}~\bibnamefont{Morigi}} \bibnamefont{and}
  \bibinfo{author}{\bibfnamefont{S.}~\bibnamefont{Fishman}},
  \bibinfo{journal}{Phys. Rev. Lett.} \textbf{\bibinfo{volume}{93}},
  \bibinfo{pages}{170602} (\bibinfo{year}{2004}).

\bibitem[{\citenamefont{Hegerfeldt and Vogt}(1990)}]{hegerfeldt90}
\bibinfo{author}{\bibfnamefont{G.~C.} \bibnamefont{Hegerfeldt}}
  \bibnamefont{and} \bibinfo{author}{\bibfnamefont{A.~W.} \bibnamefont{Vogt}},
  \bibinfo{journal}{Phys. Rev. A} \textbf{\bibinfo{volume}{41}},
  \bibinfo{pages}{2610} (\bibinfo{year}{1990}).

\bibitem[{\citenamefont{Prestage et~al.}(1991)\citenamefont{Prestage, Williams,
  Maleki, Djomehri, and Harabetian}}]{prestage91}
\bibinfo{author}{\bibfnamefont{J.~D.} \bibnamefont{Prestage}},
  \bibinfo{author}{\bibfnamefont{A.}~\bibnamefont{Williams}},
  \bibinfo{author}{\bibfnamefont{L.}~\bibnamefont{Maleki}},
  \bibinfo{author}{\bibfnamefont{M.~J.} \bibnamefont{Djomehri}},
  \bibnamefont{and}
  \bibinfo{author}{\bibfnamefont{E.}~\bibnamefont{Harabetian}},
  \bibinfo{journal}{Phys. Rev. Lett.} \textbf{\bibinfo{volume}{66}},
  \bibinfo{pages}{2964} (\bibinfo{year}{1991}).

\bibitem[{\citenamefont{Hornek\ae{}r and Drewsen}(2002)}]{hornekaer02}
\bibinfo{author}{\bibfnamefont{L.}~\bibnamefont{Hornek\ae{}r}}
  \bibnamefont{and} \bibinfo{author}{\bibfnamefont{M.}~\bibnamefont{Drewsen}},
  \bibinfo{journal}{Phys. Rev. A} \textbf{\bibinfo{volume}{66}},
  \bibinfo{pages}{013412} (\bibinfo{year}{2002}).

\bibitem[{\citenamefont{Drewsen et~al.}(1998)\citenamefont{Drewsen, Brodersen,
  Hornek\ae{}r, Hangst, and Schifffer}}]{drewsen98}
\bibinfo{author}{\bibfnamefont{M.}~\bibnamefont{Drewsen}},
  \bibinfo{author}{\bibfnamefont{C.}~\bibnamefont{Brodersen}},
  \bibinfo{author}{\bibfnamefont{L.}~\bibnamefont{Hornek\ae{}r}},
  \bibinfo{author}{\bibfnamefont{J.~S.} \bibnamefont{Hangst}},
  \bibnamefont{and} \bibinfo{author}{\bibfnamefont{J.~P.}
  \bibnamefont{Schifffer}}, \bibinfo{journal}{Phys. Rev. Lett.}
  \textbf{\bibinfo{volume}{81}}, \bibinfo{pages}{2878} (\bibinfo{year}{1998}).

\end{thebibliography}

\end{document}